\documentclass[aps,showpacs,twocolumn]{revtex4}
\usepackage{epsfig}
\usepackage{graphicx,color,dcolumn}
\usepackage{epstopdf}
\usepackage{amsmath,amssymb}
\usepackage{multirow}
\usepackage{diagbox}
\usepackage{changes}
\usepackage{booktabs}
\usepackage{threeparttable}
\usepackage{subfigure}
\usepackage{hyperref}

\begin{document}
\title{Investigating the nature of $N(1535)$ and $\Lambda(1405)$ in a quenched chiral quark model}
\author{
Yue Tan\textsuperscript{a},
Zi-Xuan Ma\textsuperscript{b},
Xiaoyun Chen\textsuperscript{c},
Xiaohuang Hu\textsuperscript{d},
Youchang Yang\textsuperscript{e},
Qi Huang\textsuperscript{b},
Jialun Ping\textsuperscript{b}
}
\email[E-mail: ]{ 181001003@njnu.edu.cn}
\email[E-mail: ]{ 19190101@njnu.edu.cn}
\email[E-mail: ]{xychen@jit.edu.cn}
\email[E-mail: ]{201001002@njnu.edu.cn }
\email[E-mail: ]{yangyc@gues.edu.cn }
\email[E-mail: ]{huangqi@nnu.edu.cn (Corresponding author) }
\email[E-mail: ]{jlping@njnu.edu.cn (Corresponding author)}
\affiliation{\textsuperscript{a}Department of Physics, Yancheng Institute of Technology, Yancheng 224000, People's Republic of China }
\affiliation{\textsuperscript{b}Department of Physics, Nanjing Normal University, Nanjing 210023, People's Republic of China }
\affiliation{\textsuperscript{c}College of Science, Jinling Institute of Technology, Nanjing 211169, People's Republic of China }
\affiliation{\textsuperscript{d}Department of Physics, Changzhou Vocational Institute of Engineering, Changzhou 213164, People's Republic of China }
\affiliation{\textsuperscript{e}Department of Physics, Guizhou University of Engineering Science, Bijie 551700,  People's Republic of China.}

\date{\today}

\begin{abstract}
In this work, we systematically study \( N(1440) \), \( N(1535) \), and \( \Lambda(1405) \) in both the quenched three-quark and five-quark frameworks using the Gaussian Expansion Method (GEM) within the chiral quark model. Our calculations show that \( N(1535) \) can be reproduced as a three-quark state (\( N(1P) \)), while \( N(1440) \) and \( \Lambda(1405) \) cannot be accommodated as the three-quark candidates, (\( N(2S) \) and \( \Lambda(1P) \)), respectively. In the five-quark framework, we find that the \( \Lambda K \) state for \( N(1535) \) can not form a bound state, while in the \( N\bar{K} \) channel there will \( \Lambda(1405) \) form a shallow bound state. Based on the complex-scaling method, we performed complete coupled-channels calculations and obtained six resonance states with energies ranging from 1.8 GeV to 2.2 GeV, in addition with one bound state located around \( \Sigma \pi \) channel. However, neither molecular candidates in \( \Lambda K \) channel for \( N(1535) \) nor \( N\bar{K} \) for \( \Lambda(1405) \) are included in these states. This is because the strong coupling between \( N\bar{K} \) and \( \Sigma \pi \) will make the \( N\bar{K} \) unbound, while the weak coupling between \( \Lambda K \) and \( \Sigma K \) can not help form a stable structure around \( \Lambda K \) threshold. Thus, under the quenched quark model, our results support \( N(1535) \) as a three-quark state, while \( N(1440) \) is neither a three-quark nor a five-quark state. In addition, we find that although \( \Lambda(1405) \) can be primarily a five-quark state, it requires a mixture of three-quark and five-quark components for stability. In the future, an exploration on the mixing effects between bare baryons with these relevant two-body hadronic channel components will be carried out to further test our conclusions.

\end{abstract}

\maketitle

\section{Introduction}

The mass reverse problem of the Roper resonances ($N(1440)$ and $N(1535)$) has long been an important topic in hadron physics \cite{Roper:1964zza}. According to the traditional quark model, $N(1440)$ is often treated as the first radial excitation of the proton ($qqq$), i.e., the $N(2S)$ state, while $N(1535)$ is usually seen as the orbital excitation of the proton ($qqq$), i.e., $N(1P)$. However, based on the energy relation $E = \hbar \omega (2n + l)$, although the energy of $N(2S)$ should be greater than that of $N(1P)$ theoretically, but experiments observed such mass reverse phenomenon. In addition, due to the heavier mass of constituent strange quark, the $\Lambda(1405)$ state, as the orbital excitation of the $\Lambda$ baryon ($qqs$), should have an energy greater than that of $N(1P)$, yet the experimental observations also show the opposite measurements. These mass reverse problems present a significant challenge to the traditional quark model, thus, investigating the internal structure of these particles is crucial for enhancing our understanding of quark model itself, and the non-perturbative effect of Quantum Chromodynamics (QCD).

Current theoretical works on Roper resonances can be mainly categorized into two types: the first type employs a modified quark model to contain the Roper resonance still into the three-quark framework \cite{Capstick:1985xss,Capstick:1992th,Isgur:1978xj,Glozman:1999vd,Garcilazo:2001md,Yang:2017qan,Ferretti:2011zz}, while the others try to interpret the it as a baryon-meson molecular structure \cite{Liu:2005pm,Kaiser:1995cy,Nieves:2001wt,Inoue:2001ip,Hyodo:2003qa,Oller:2000ma,Bruns:2010sv,Khemchandani:2013nma,Sekihara:2015gvw,Molina:2023jov,Oller:2000fj,Jido:2003cb,Mai:2014xna,Qin:2020gxr,BaryonScatteringBaSc:2023ori,Guo:2023wes}. Based on a relativized quark model that incorporates confinement and one-gluon exchange (OGE) interactions, S. Capstick and N. Isgur systematically studied the S-wave and P-wave baryons in both light and heavy sectors. Their results show that although the $N(1535)$, $N(1440)$, and $\Lambda(1405)$ states can be accommodated within the three-quark framework \cite{Capstick:1985xss,Capstick:1992th,Isgur:1978xj}, the energy of $N(2S)$ is still approximately 100 MeV higher than that of $N(1440)$ \cite{Capstick:1992th}. Then, Glozman et al. emphasized the importance of Goldstone exchange potential. Based on this, they provided a reasonable explanation for the mass inversion problem of the Roper resonances ($N(1535)$ and $N(1440)$), although there remains some discrepancy with experimental values \cite{Glozman:1999vd}. In addition, they found that the energy of $\Lambda(1P)$ is still significantly higher than the experimental value of $\Lambda(1405)$, which is around 1.5 GeV. After that, Ref. \cite{Garcilazo:2001md} applied the Faddeev approach and incorporated the pseudoscalar and scalar Goldstone boson exchange potentials into their model. They concluded that the resulting spectrum is reasonable and in good agreement with the $NN$ phenomenology, especially when the standard one-gluon exchange force is included. Similarly, Ref.  \cite{Yang:2017qan} applied the chiral quark model and included contributions from scalar and pseudoscalar meson exchanges. Within a certain error margin, they addressed the mass inversion problem of the Roper resonance. However, the energy of $\Lambda(1P)$ remains significantly higher than that of $\Lambda(1405)$. After employing the relativistic quark-diquark mass operator, Ferretti et al. \cite{Ferretti:2011zz} achieved that the energy of $N(2S)$ to be lower than that of $N(1P)$. However, in their work, the mass of $N(2S)$ is calculated around 1.5 GeV, which made it not a suitable candidate for $N(1440)$. Hence, until now, there still no model that can explain $N(1535)$, $N(1440)$, and $\Lambda(1405)$ simultaneously well as experimental observations as traditional baryon states.

Regarding the mass inverse problem in the Roper resonance, B. S. Zou et al. \cite{Liu:2005pm} argue that $N(1440)$ and $N(1535)$ are not purely three-quark states, but rather contain significant five-quark components ($qqqq\bar{q}$ and $qqq s \bar{s}$). Actually, earlier in Ref.\cite{Kaiser:1995cy}, after using the effective chiral lagrangian at next-to-leading order, the authors already obtained a quasi-bound $K\Sigma$-$K\Lambda$ state, whose mass indicated that it might be a good candidate of $N(1535)$. Similarly, after considering chiral symmetry, the coupled-channel Bethe-Salpeter equation was applied to study the coupling among four physical channels ($\pi N$, $\eta N$, $K\Lambda$, and $K\Sigma$), then resonances seemed to be related with $N(1535)$ and $N(1650)$ were also obtained \cite{Nieves:2001wt}. Apart from two-body channels, Ref. \cite{Inoue:2001ip} demonstrated that the $\pi\pi N$ channel can also significantly affect the results, after considering this three-body channel, the mass, decay width, and branching ratios of $N(1535)$ were in good agreement with experimental measurements, and such conclusion can also be similarly found in Ref.~\cite{Hyodo:2003qa}.

To explain the generation mechanisms of Roper resonances as nucleon-like pentaquark states, studies on meson-baryon interactions were carried on. Under the framework of unitarized coupled-channel chiral perturbation theory \cite{Oller:2000ma}, $N(1535)$ could be generated through strong channel couplings, and $K\Sigma$-$K\Lambda$ component was found to be dominant in its wave function, which means conclusions in Refs. \cite{Kaiser:1995cy,Nieves:2001wt} got supported. Then, by studying $S$-wave pion-nucleon scattering within the unitarized chiral effective Lagrangian that includes all possible two-body contact terms, Ref.\cite{Bruns:2010sv} obtained two states that can relate to $N(1535)$ and $N(1635)$, which was then supported by the work of Ref. \cite{Khemchandani:2013nma}. While in Ref. \cite{Sekihara:2015gvw}, within the framework of chiral perturbation theory up to next-to-leading order, the scattering amplitudes of the $N(1535)$ and $N(1650)$ resonances were investigated, and it was found that the $\pi N$, $\eta N$, $K \Lambda$, and $K \Sigma$ components contribute negligibly to these resonances \cite{Sekihara:2015gvw}. Recently, correlation functions have also been used to investigate the mass of $N(1535)$. By fitting the data within a general framework, Ref.~\cite{Molina:2023jov} predicted the existence of a bound state with isospin $I = 1/2$, whose mass is just around the region of $N^(1535)$.

As for $\Lambda(1405)$, current problems in containg it into traditional baryon spectra also give birth on the idea that if $\Lambda(1405)$ may also be a molecular state. For example, in Ref.~\cite{Oller:2000fj,Jido:2003cb,Mai:2014xna}, the authors studid the S-wave kaon-nucleon interactions with strangeness $S = -1$ by using a novel relativistic chiral unitary approach based on coupled channels, and they obtained two pole structures as two bound states simultaneously under $\bar{K} N$ and $\Sigma \pi$ coupled channels, and such conclusion were also supported by the quark model calculation \cite{Qin:2020gxr}. The latest lattice results showed that, after studying the coupled-channel $\bar{K} N$-$\Sigma \pi$ scattering amplitudes, a virtual state below the $\pi\Sigma$ threshold and a resonance pole just below the $\bar{K}N$ threshold were obtained \cite{BaryonScatteringBaSc:2023ori}. Very recently, the authors \cite{Guo:2023wes} systematically studied the pole position behaviors of $\Lambda(1380)$, $\Lambda(1405)$, and $\Lambda(1680)$ after considering the evolution of symmetry from SU(3) limit to SU(3) broken, and found that the symmetry of two octets caused by the leading order Weinberg-Tomozawa term will be broken by the next-to-leading order term, and the state related to $\Lambda(1405)$ has always two different poles for any SU(3) limit, one in singlet and the other in octet, while the $\Lambda(1680)$ is degenerate with the heavier pole at leading order in that limit.

Thus, based on the above works, in the current community, $N(1535)$ can be interpreted as either a three-quark or a five-quark state, while $\Lambda(1405)$ is more likely to be a $N\bar{K}$ molecular state, possibly even one of the two-pole structures that come from $N\bar{K}$-$\Sigma \pi$ coupled channel effect. As for $N(1440)$, forcibly accommodating it within a three-quark structure will more or less trigger unavoidable problems, if one wants to explain the whole nucleon family well. Thus, to further understand the nature of Roper resonances, studies on the interactions between bare baryon states with relevant scattering channels may be necessary, i.e., by using the language of quark model, the unquenched effect should be considered. However, the complexity of few-body problem itself tells that, taking a first look on the behaviors of spectra under the quenched quark model is always helpful, which can at least tell us the basic properties of these states. Thus, to address the problem of Roper resonance, as our first step, we use the Rayleigh-Ritz variational method combined with the Gaussian expansion method (GEM) to study the mass problem of $N(1440)$, $N(1535)$, and $\Lambda(1405)$ in the three-quark as well as five-quark framework. In addition, to locate possibly existed resonance states, we incorporate the complex-scaling method in our calculations.

The structure of this paper is as follows: After the introduction, Section II provides a brief description of the quark model, wave functions, and an introduction on the complex-scaling method. Then it follows our results and discussion in Section III. Finally, this paper ends up with a summary.

\section{Model setup} \label{wavefunction and chiral quark model}
\subsection{Chiral quark model}

In our model, the dynamics of the Roper resonances ($N(1535)$, $N(1440)$) and the $\Lambda(1405)$ are described via Gauss Expansion Method based on the chiral quark model, whose Hamiltonian includes mass terms, kinetic terms, and potential terms as
\begin{eqnarray}
H &=&\sum_{i=1}^n ( m_i +\frac{\vec{p}_{i}^{~2}}{2 m_{i}} ) -T_{c} +  \sum_{i<j=1}^n V(r_{ij}) ,
\end{eqnarray}
where $m_i$ denotes the quark mass, $\vec{p}_i$ represents the quark momentum, $T_c$ is the center-of-mass kinetic energy of the quark system, and $V(r_{ij})$ represents the potential term. In the Jacobi coordinate system, for a three-quark system, the kinetic term $\sum_{i=1}^n \left( m_i + \frac{\vec{p}_{i}^{~2}}{2 m_i} \right) - T_c$ can be reduced to
\begin{eqnarray}
\label{T3}
\sum_{i=1}^3 ( m_i +\frac{\vec{p}_{i}^{~2}}{2 m_{i}} ) -T_{c} = \frac{\vec{p}_{12}^{~2}}{2\mu_{12}} + \frac{\vec{p}_{12,3}^{~2}}{2\mu_{12,3}},
\end{eqnarray}
with $p_{i,j}$ and $\mu_{i,j}$ are relative momentum and reduced mass, respectively. While for a five-quark system, the kinetic term will be simplified as
\begin{eqnarray}
\label{T5}
&&\sum_{i=1}^5 ( m_i +\frac{\vec{p}_{i}^{~2}}{2 m_{i}} ) -T_{c}\nonumber\\ &&\qquad = \frac{\vec{p}_{12}^{~2}}{2\mu_{12}} + \frac{\vec{p}_{12,3}^{~2}}{2\mu_{12,3}} + \frac{\vec{p}_{45}^{~2}}{2\mu_{45}} + \frac{\vec{p}_{123,45}^{~2}}{2\mu_{123,45}}.
\end{eqnarray}
And in Eq.\ref{T3} and Eq.\ref{T5}, the specific forms of relative momenta and reduced masses can be written explicitly as
\begin{subequations}
\label{moment}
\begin{align}
\mu_{12}&=\frac{m_{{1}}  m_{{2}}}{m_{{1}} + m_{{2}}}, \mu_{45}=\frac{m_{{4}}  m_{{5}}}{m_{{4}} + m_{{5}}},   \\
\mu_{12,3}&=\frac{(m_1+m_2) m_3}{m_1+m_2+m_3},\\
\mu_{123,45}&=\frac{(m_1+m_2+m_3) (m_4+m_5)}{m_1+m_2+m_3+m_4+m_5},\\
\vec{p}_{12}&=\frac{m_2\vec{p}_1-m_1\vec{p}_2}{m_1+m_2},\vec{p}_{45}=\frac{m_5\vec{p}_4-m_4\vec{p}_5}{m_4+m_5},  \\
\vec{p}_{12,3}&=\frac{m_3\vec{p}_{12}-(m_1+m_2)\vec{p}_{3}}{m_1+m_2+m_3},  \\
\vec{p}_{123,45}&=\frac{(m_4+m_5)\vec{p}_{123}-(m_1+m_2+m_3)\vec{p}_{45}}{m_1+m_2+m_3+m_4+m_5}.
\end{align}
\end{subequations}

Then, for potential terms, in light quark systems, the spontaneous breaking of chiral symmetry may play a significant role. Therefore, for the potential term $V(r_{ij})$, our model not only includes the confinement potential ($V_{con}(r_{ij})$) and the one-gluon exchange potential ($V_{oge}(r_{ij})$), but also incorporates the Goldstone boson exchange potential ($V_{\chi}(r_{ij})$) and scalar boson exchange potential ($V_{s}(r_{ij})$). The potential used in this work are totally written as
\begin{eqnarray}
V(r_{ij}) &=& \sum_{i<j=1}^n \left[ V_{con}(r_{ij})+V_{oge}(r_{ij}) + \sum_{s=\sigma} V_{s}(r_{ij})\right. \nonumber\\
          &&\left. + \sum_{\chi=\pi,\eta,K} V_{\chi}(r_{ij})\right].
\end{eqnarray}

For the confinement potential $V_{con}(r_{ij})$, in this work, we adopt the following quadratic form
\begin{align}
    V_{con}(r_{ij}) &= \left( -a_{c} r_{ij}^{2} - \Delta \right) \boldsymbol{\lambda}_i^c \cdot \boldsymbol{\lambda}_j^c, \\
    \nonumber
\end{align}
where $\boldsymbol{\lambda}^{c}$ are $SU(3)$ color Gell-Mann matrices, $a_{c}$, $\Delta$ are model parameters. For this quadratic form choice on confinement, we want to emphasize here that there are two reasons, one is that for ground and low-lying states, unquenched effect is assumed to be small, thus using an unquenched confinement under this case may be a good approximation. The other is that our next step is to systematically study the unquenched effect, choosing this kind of confinement possibly may let us in the future just need to do a fine-tuning on model parameters, which can be seen as a consistency between this and future works.

Then, the one-gluon exchange potential $V_{oge}(r_{ij})$ consists of two parts, the Coulomb term and the color-magnetic term, which reads
\begin{eqnarray}
 V_{oge} (r_{ij}) &=& \frac{\alpha_s}{4} \boldsymbol{\lambda}_i^c \cdot \boldsymbol{\lambda}_j^c \left[\frac{1}{r_{ij}} - \frac{2}{3m_i m_j} \hat{S}_i \cdot \hat{S}_j\right.\nonumber\\
 &&\times\left.\frac{e^{-r_{ij}/r_0(\mu_{ij})}}{r_{ij} r_0^2(\mu_{ij})} \right],
\end{eqnarray}
where $\boldsymbol{\lambda}^{c}$ denotes the $SU(3)$ Gell-Mann matrices acting on the color wave functions of the quark system, $r_0$ is a model parameter, $\alpha_s$ is the coupling constant determined from experimental fitting values, and $\hat{S}_i$ represents the spin operator acting on the spin-$\frac{1}{2}$ wave functions of the quark system.

Finally, the explicit forms of the potentials that describing the Goldstone boson exchange, which is the most significant feature of the chiral quark model, is that
\begin{eqnarray}
V_{\pi}(r_{ij}) &=& \frac{g_{ch}^2}{4\pi} \frac{m_{\pi}^2}{3m_im_j} \frac{\Lambda_{\pi}^2 m_\pi}{\Lambda_{\pi}^2 - m_{\pi}^2} \hat{S}_i \cdot \hat{S}_j \sum_{a=1}^3 \lambda_i^a \lambda_j^a \nonumber\\
&&\times \left[ Y(m_\pi r_{ij}) - \frac{\Lambda_{\pi}^3}{m_{\pi}^3} Y(\Lambda_{\pi} r_{ij}) \right], \nonumber \\
V_{K}(r_{ij}) &=& \frac{g_{ch}^2}{4\pi} \frac{m_{K}^2}{3m_im_j} \frac{\Lambda_K^2 m_K}{\Lambda_K^2 - m_K^2}  \hat{S}_i \cdot \hat{S}_j \sum_{a=1}^3 \lambda_i^a \lambda_j^a\nonumber\\
&&\times \left[ Y(m_K r_{ij}) - \frac{\Lambda_K^3}{m_K^3} Y(\Lambda_K r_{ij}) \right], \\
V_{\eta}(r_{ij}) &=& \frac{g_{ch}^2}{4\pi} \frac{m_{\eta}^2}{3m_im_j} \frac{\Lambda_{\eta}^2 m_{\eta}}{\Lambda_{\eta}^2 - m_{\eta}^2} \hat{S}_i \cdot \hat{S}_j \left( \lambda_i^8 \lambda_j^8 \cos\theta_P \right.\nonumber\\
&&\left. - \lambda_i^0 \lambda_j^0 \sin \theta_P \right)\left[ Y(m_\eta r_{ij}) - \frac{\Lambda_{\eta}^3}{m_{\eta}^3} Y(\Lambda_{\eta} r_{ij}) \right], \nonumber \\
V_{\sigma}(r_{ij}) &=& -\frac{g_{ch}^2}{4\pi} \frac{\Lambda_{\sigma}^2 m_\sigma}{\Lambda_{\sigma}^2 - m_{\sigma}^2} \left[ Y(m_\sigma r_{ij}) - \frac{\Lambda_{\sigma}}{m_\sigma} Y(\Lambda_{\sigma} r_{ij}) \right], \nonumber
\end{eqnarray}
where $\boldsymbol{\lambda}^{a}$ denotes the $SU(3)$ Gell-Mann matrices acting on the flavor wave functions of the quark system. $Y(x)$ is the Yukawa function explicitly given by $Y(x)=\frac{e^{-x}}{x}$. $\Lambda_{\chi}$ serves as the cut-off parameter, and $g^2_{ch}/4\pi$ denotes the Goldstone-quark coupling constant. Here, the masses of Goldstone bosons $\pi$, $K$, and $\eta$ are denoted by $m_{\pi}$, $m_{\eta}$, and $m_{K}$, respectively, while $m_{\sigma}$ is determined by the relation
\begin{align}
m_{\sigma}^2 \approx m_{\pi}^2+4 m_{u,d}^2.
\end{align}
\begin{table*}[tp]
  \centering
  \fontsize{9}{8}\selectfont
  \makebox[\textwidth][c]{
   \begin{threeparttable}
   \caption{Results of the hadron spectrum calculation.\label{fitresults}}
    \begin{tabular}{cccccccccccccc}
    \hline    \hline
                                              & $\pi$  & $\eta$ & $\rho$ & $\omega$ & $\bar{K}(K)$  & $\bar{K}^{*}(K^{*})$ & N    & $\Lambda$  &$\Sigma$  &$\Sigma^{*}$ &$\Xi$ &$\Xi^{*}$  \\
    This work                                 &  143   &  598   & 785    &   798    &   495         &   913                & 939  &   1071     &  1215    &   1345&   1369 &   1479  \\
    EXP.(PDG)\cite{ParticleDataGroup:2024cfk} & 139.57 ~~& 547.86 ~~& 775.26 ~~&  782.66  ~~& 497.61 ~~&   895.55   ~~  & 939.5~~&  1115.6    ~&  1192.6 ~ & 1383.7 ~& 1314.8 ~&1531.8 \\
\hline \hline
    \end{tabular}
   \end{threeparttable}}
  \end{table*}

After fitting the ground states of light mesons and baryons, all the model parameters are determined, which are collected into Table~\ref{modelparameters},
\begin{table}[tbp]
\begin{center}
\caption{Quark model parameters ($m_{\pi}=0.7$ $fm^{-1}$, $m_{\sigma}=3.42$ $fm^{-1}$, $m_{\eta}=2.77$ $fm^{-1}$, $m_{K}=2.51$ $fm^{-1}$).\label{modelparameters}}
\begin{tabular}{cccc}
\hline\hline\noalign{\smallskip}
Quark masses   &$m_u=m_d$(MeV)     &490  \\
               &$m_{s}$(MeV)         &511  \\

Goldstone bosons
                   &$\Lambda_{\pi}(fm^{-1})$     &3.5  \\
                   &$\Lambda_{\eta}(fm^{-1})$     &2.2  \\
                   &$\Lambda_{\sigma}(fm^{-1})$     &7.0  \\
                   &$\Lambda_{a_0}(fm^{-1})$     &2.5  \\
                   &$\Lambda_{f_0}(fm^{-1})$     &1.2  \\
                   &$g_{ch}^2/(4\pi)$                &0.54  \\
                   &$\theta_p(^\circ)$                &-15 \\

Confinement             &$a_{c}$ (MeV$\cdot fm^{-2}$)     &98 \\
                        &$\Delta_{qq/q\bar{q}}$(MeV)       &-91.1/-10.1 \\
                        &$\Delta_{qs/q\bar{s}}$(MeV)       &-58.4/-10.0 \\
                        &$\Delta_{s\bar{s}}$(MeV)       &-18.1 \\

OGE                 & $\alpha_{qq/q\bar{q}}$        &0.69/1.34 \\
                    & $\alpha_{qs/q\bar{s}}$        &0.90/1.15 \\
                    & $\alpha_{s\bar{s}}$        &0.91 \\
                    &$\hat{r}_0$(MeV)    &80.9 \\
\hline\hline
\end{tabular}
\end{center}
\end{table}
while the fit results are presented in Table~\ref{fitresults}.

\subsection{The wave function of $N(1440)$, $N(1535)$ and $\Lambda(1405)$}
After introducing the Hamiltonian and model parameters, next, we present our constructions on the wave functions of $N(1440)$, $N(1535)$, and $\Lambda(1405)$ from both the three-quark and five-quark perspectives. Since in this work, for five-quark perspective, we focus on their molecular nature, therefore, we first construct the color singlet three-quark and two-quark wave functions. Then, following the principle of group theory, we couple the three-quark and two-quark wave functions to form the five-quark wave function. Since quarks have four degrees of freedom, namely flavor ($\psi$), orbital ($\phi$), spin ($\chi$), and color ($\xi$), their specific construction processes will be presented respectively as follows.

Firstly, we consider the flavor part. For the $N(1440)$ and $N(1535)$, traditional baryonic state explanations on them are based on the $qqq$ configuration, while for the $\Lambda(1405)$, the three-quark explanation corresponds to the $qqs$ configuration. For their pentaquark configurations, considering the possible thresholds and the distinctiveness of the $\sigma$ meson, it may be unlikely that $N(1440)$ is a very good five-quark molecular state ($N\sigma$), so we will not discuss it further here but plan to collect it into our future works on unquenched quark model. On the other hand, $N(1535)$ is commonly interpreted as a $qqs$-$\bar{s}q$ configuration, and $\Lambda(1405)$ corresponds to the $qqq$-$\bar{q}s$ or $qqs$-$\bar{q}q$ configuration.

For $N(1440)$ and $N(1535)$, they have isospin $I = \frac{1}{2}$. Since isospin is a good quantum number, according to Wigner-Eckart theorem, on the spectrum side we will only consider the case where the third component of isospin is $I_z = \frac{1}{2}$, which gives that
\begin{eqnarray}
\psi_{\frac{1}{2},\frac{1}{2}}^{B_1} &=& \frac{1}{\sqrt{2}}(udu - duu), \\
\psi_{\frac{1}{2},\frac{1}{2}}^{B_2} &=& \frac{1}{\sqrt{6}}(2uud - udu - duu).
\end{eqnarray}
Here, the wave function $\psi_{\frac{1}{2},\frac{1}{2}}^{B_1}$ is the flavor-antisymmetric wave function, and $\psi_{\frac{1}{2},\frac{1}{2}}^{B_2}$ is the flavor-symmetric wave function. For the five-quark interpretation of $N^\ast$, its structure is $qqs$-$\bar{s}q$. The corresponding flavor wave functions for these structures are as follows
\begin{eqnarray}
\psi_{\frac{1}{2},\frac{1}{2}}^{P_1} &=& \frac{1}{\sqrt{2}}(uds\bar{s}u - dus\bar{s}u), \\
\psi_{\frac{1}{2},\frac{1}{2}}^{P_2} &=& \frac{1}{\sqrt{6}}(2uus\bar{s}d - uds\bar{s}u - dus\bar{s}u).
\end{eqnarray}

For the $\Lambda(1405)$, it has isospin $I = 0$ and its third component can only take as $I_z = 0$. In the three-quark framework, considering the explicitly breaking of SU(3) symmetry, its flavor wave function only has flavor-antisymmetric part as
\begin{eqnarray}
\psi_{0,0}^{B_3} = \frac{1}{\sqrt{2}}( uds - dus).
\end{eqnarray}
On the five-quark prospect, there are two possible molecular configurations: $qqq$-$\bar{q}s$ and $qqs$-$\bar{q}q$. For the $qqq$-$\bar{q}s$ configuration, the wave functions are
\begin{eqnarray}
\psi_{0,0}^{P_5} &=& \frac{1}{\sqrt{4}}(udd\bar{d}s - dud\bar{d}s + udu\bar{u}s - duu\bar{u}s), \\
\psi_{0,0}^{P_6} &=& \frac{1}{\sqrt{12}}(2ddu\bar{d}s - udd\bar{d}s - dud\bar{d}s - 2uud\bar{u}s \nonumber\\
                 &&+ udu\bar{u}s + duu\bar{u}s),
\end{eqnarray}
where $\psi_{0,0}^{P_5}$ corresponds to flavor-antisymmetric and $\psi_{0,0}^{P_6}$ corresponds to flavor-symmetric. While for the $qqs$-$\bar{q}q$ configuration, the corresponding wave functions are
\begin{eqnarray}
\psi_{0,0}^{P_7} &=& \frac{1}{\sqrt{4}}(uds\bar{u}u + uds\bar{d}d - dus\bar{u}u - dus\bar{d}d), \\
\psi_{0,0}^{P_8} &=& \frac{1}{\sqrt{12}}(2dds\bar{d}u + uds\bar{u}u - uds\bar{d}d + dus\bar{u}u \nonumber\\
&&- dus\bar{d}d - 2uus\bar{u}d).
\end{eqnarray}

Next, for the orbital part, under the Jacobian coordinate system, the three-quark system has two relative motions, while the five-quark system has four relative motions. For each relative motion, the radial wave function is expanded through the Gaussian-Expansion Method (GEM), in which each radial wave function is given by
\begin{align}
\phi_{nlm}(r)  = N_{nl}r^{l} e^{-\nu_{n}r^2}Y_{lm}(r),
\end{align}
with $N_{nl}$ being the normalization constants as
\begin{align}
N_{nl}=\left[\frac{2^{l+2}(2\nu_{n})^{l+\frac{3}{2}}}{\sqrt{\pi}(2l+1)}
\right]^\frac{1}{2}.
\end{align}

Hereafter, we will abbreviate $\phi_{nlm}(r)$ as $\phi_{l}(r)$. Then, the general expressions for the three-quark orbital wave function $\phi_{L,m_L}^{B}(r)$ and the five-quark wave function $\phi_{L,m_L}^{P}(r)$ can be written as
\begin{eqnarray}
\phi_{L,m_L}^{B}(r) &=& \phi_{l_{12}}(r_{12}) \phi_{l_3}(r_3), \\
\phi_{L,m_L}^{P}(r) &=& \phi_{l_{12}}(r_{12}) \phi_{l_3}(r_3) \phi_{l_{45}}(r_{45}) \phi_{l_{123,45}}(R).
\end{eqnarray}
Here, $r_{12}$ and $l_{12}$ represent the relative motion and relative orbital angular momentum between the quarks with labels 1 and 2 in the three-quark system, respectively, while $r_3$ and $l_3$ represent the relative motion and relative angular momentum between the quark labeled 3 and the (12)-di-quark. For the five-quark molecular states, we can regard the system as a coupling of the orbital wave functions of a three-quark cluster and a two-quark cluster. Specifically, $r_{45}$ and $l_{45}$ represent the relative motion and relative orbital angular momentum between the quarks labeled 4 and 5 in the two-quark cluster, while $R$ and $l_{123,45}$ represent the relative motion and relative orbital angular momentum between the baryon and meson clusters. Since in this work, for the three-quark system, we only consider $S$-wave and $P$-wave wave functions, and for the five-quark system, we only consider the $S$-wave case, thus, all the coupling coefficient between the orbital wave functions is equal to 1.

For the spin of the three-quark system, there are two possible quantum numbers, $\frac{1}{2}$ and $\frac{3}{2}$. The state with $\frac{1}{2}$ can arise either from a combination of $0 \otimes \frac{1}{2}$ or $1 \otimes \frac{1}{2}$. The former is labeled as $\chi_{\frac{1}{2}}^{B_1}$, corresponding to the antisymmetric relation between particles 1 and 2, while the latter is labeled as $\chi_{\frac{1}{2}}^{B_2}$, corresponding to the symmetric relation between particles 1 and 2. Considering the third component of spin, the three-quark spin wave function with quantum number $\frac{1}{2}$ has a total of four wave functions, which are
\begin{eqnarray}
\chi_{\frac{1}{2},\frac{1}{2}}^{B_1} &=& \frac{1}{\sqrt{2}} \left( \alpha \beta \alpha - \beta \alpha \alpha \right), \\
\chi_{\frac{1}{2},-\frac{1}{2}}^{B_1} &=& \frac{1}{\sqrt{2}} \left( \alpha \beta \beta - \beta \alpha \beta \right), \\
\chi_{\frac{1}{2},\frac{1}{2}}^{B_2} &=& \frac{1}{\sqrt{6}} \left( 2 \alpha \alpha \beta - \alpha \beta \alpha - \beta \alpha \alpha \right), \\
\chi_{\frac{1}{2},-\frac{1}{2}}^{B_2} &=& \frac{1}{\sqrt{6}} \left( \alpha \beta \beta + \beta \alpha \beta - 2 \beta \beta \alpha \right).
\end{eqnarray}

Similarly, the total spin wave function with quantum number $\frac{3}{2}$ has four components, resulting in four wave functions. All of them are symmetric with respect to particles 1 and 2, whose explicit expressions are as follows
\begin{eqnarray}
\chi_{\frac{3}{2},\frac{3}{2}}^{B_3} &=& \alpha \alpha \alpha, \\
\chi_{\frac{3}{2},\frac{1}{2}}^{B_3} &=& \frac{1}{\sqrt{3}} (\alpha \alpha \beta + \alpha \beta \alpha + \beta \alpha \alpha), \\
\chi_{\frac{3}{2},-\frac{1}{2}}^{B_3} &=& \frac{1}{\sqrt{3}} (\alpha \beta \beta + \beta \alpha \beta + \beta \beta \alpha), \\
\chi_{\frac{3}{2},-\frac{3}{2}}^{B_3} &=& \beta \beta \beta.
\end{eqnarray}

On the pentaquark side, for both the $N(1535)$ and the $\Lambda(1405)$, the total spin is $\frac{1}{2}$. This can be interpreted as the coupling between a three-quark spin wave function with $S = \frac{1}{2}$ or $S = \frac{3}{2}$ and a two-quark system with spin $S = 0$ or $S = 1$. Based on the previous discussion on the three-quark spin wave functions, the spin wave function for the three-quark system with spin $\frac{1}{2}$ can be either symmetric or antisymmetric. This wave function can couple with both $S = 0$ and $S = 1$ to produce the total five-quark spin wave function with spin $\frac{1}{2}$, denoted as $\chi_{\frac{1}{2}, \frac{1}{2}}^{P_i}$. Therefore, there are four possible cases
\begin{eqnarray*}
\chi_{\frac{1}{2}, \frac{1}{2}}^{P_1} &=& \frac{1}{2} ( \alpha \beta \alpha \beta - \alpha \beta \alpha \beta \alpha - \beta \alpha \alpha \alpha + \beta \alpha \alpha \beta), \\
\chi_{\frac{1}{2}, \frac{1}{2}}^{P_2} &=& \frac{1}{2\sqrt{3}} ( 2 \alpha \beta \alpha \beta \alpha - 2 \alpha \beta \beta \alpha \alpha - \alpha \beta \alpha \beta + \alpha \beta \alpha \beta \alpha  \\
&&- \beta \alpha \alpha \alpha + \beta \alpha \beta \alpha), \\
\chi_{\frac{1}{2}, \frac{1}{2}}^{P_3} &=& \frac{1}{2\sqrt{3}} ( 2 \alpha \beta \beta \alpha \alpha - \alpha \beta \alpha \beta \alpha - \alpha \beta \alpha \beta + \alpha \beta \alpha \beta \alpha \\
&&+ \beta \alpha \alpha \alpha + \beta \alpha \beta \alpha), \\
\chi_{\frac{1}{2}, \frac{1}{2}}^{P_4} &=& \frac{1}{6} ( 2 \alpha \beta \beta \alpha \alpha + 2 \beta \alpha \alpha \beta \alpha - 4 \beta \beta \alpha \alpha \alpha - 2 \alpha \beta \alpha \beta \alpha\\
&&- 2 \alpha \beta \beta \alpha \alpha + \alpha \beta \alpha \alpha \beta+ \alpha \beta \alpha \beta \alpha \beta\nonumber\\
&&+ \beta \alpha \alpha \alpha + \beta \alpha \alpha \beta \alpha).
\end{eqnarray*}

The other possible coupling scheme for the total spin of the five-quark system of $N(1535)$ and $\Lambda(1405)$ is the case where the three-quark system has spin $S = \frac{3}{2}$ and the two-quark system has spin $S = 1$. The wave function $\chi_{\frac{1}{2}, \frac{1}{2}}^{P_5}$ can be written as
\begin{eqnarray*}
\chi_{\frac{1}{2}, \frac{1}{2}}^{P_5} &=& \frac{1}{3\sqrt{2}} ( 3 \alpha \alpha \beta \beta \beta + \alpha \alpha \beta \alpha \beta + \alpha \beta \alpha \beta \alpha + \beta \beta \alpha \alpha \alpha  \\
&&- \alpha \alpha \beta \alpha \alpha - \alpha \beta \alpha \alpha \beta - \beta \alpha \beta \alpha \alpha - \alpha \alpha \beta \beta \alpha \\
&&- \alpha \beta \alpha \beta \alpha - \beta \alpha \beta \alpha \beta).
\end{eqnarray*}

For the color wave function, the three-quark system must be in a color-neutral state. Therefore, it can be written as
\begin{eqnarray}
\xi^{B} &=& \frac{1}{\sqrt{6}} ( \text{rgb} -{grb} +{gbr} -{brg} +{bgr} ).
\end{eqnarray}
Then, under molecular state configuration, the color wave function of the five-quark system can be viewed as the coupling of the color wave functions of the three-quark system and the two-quark system. Thus, It can be obtained by coupling two color singlet wave functions, denoted by \(\xi^{P_1}\), as
\begin{eqnarray}
\xi^{P_1} &=& \frac{1}{\sqrt{6}} ({rgb} -{rbg} +{gbr} -{grb} +{brg} -{bgr} ) \\ \nonumber
&&\times \frac{1}{\sqrt{3}} ( \bar{r}r + \bar{g}g + \bar{b}b )
\end{eqnarray}

Finally, the total wave function is the tensor product of the above four components
\begin{eqnarray}
\Psi_{J,mJ}^{B_{ijkl}}(r) &=& \mathcal{A}_3 \left[ \phi_l^{B_i} \chi^{B_j} \right]_{J,mJ} \psi_{I,mI}^{B_k} \xi^{B_l} \\
\Psi_{J,mJ}^{P_{ijkl}}(r) &=& \mathcal{A}_5 \left[ \phi_l^{P_i} \chi^{P_j} \right]_{J,mJ} \psi_{I,mI}^{P_k} \xi^{P_l},
\end{eqnarray}
where \( \mathcal{A}_3 \) is the antisymmetrization operatorfor the three-quark system, and \( \mathcal{A}_5 \) is the antisymmetrization operatorfor the five-quark system.

\subsection{Complex-Scaling Method}

The complex-scaling method is a powerful technique for locating resonant states, originally introduced in Ref. \cite{Aguilar:1971ve,Balslev:1971vb}. In this method, all  coordinates \( \vec{r} \) in the Hamiltonian \( H \) are replaced by \( \vec{r} e^{i\theta} \), where \( \theta \) is a complex scaling factor. By solving the Schrodinger equation in the complex plane, both energy and width (i.e., the decay rate of the resonance) information of resonant states can be simultaneously obtained.

In the complex-scaling representation, energy $M$ is plotted along the horizontal axis, and the half-width \( \Gamma / 2 \) is plotted along the vertical axis. As the scaling angle \( \theta \) varies, the system exhibits the following different behaviors.

\begin{itemize}
    \item If the system contains bound states, the corresponding points representing bound states will converge to the real axis.
    \item The points corresponding to scattering states will lie along the same line, i.e., all these points having the same \( \theta \) value.
    \item The points representing resonances do not lie on the scattering line, but remain invariant as \( \theta \) changes. The vertical coordinate of these points corresponds to the half-width of the resonance, i.e., \( \Gamma / 2 \).
\end{itemize}

\begin{figure}[htbp]
    \centering
    \includegraphics[width=0.5\textwidth, trim=4.5cm 5.0cm 4.5cm 2.0cm, clip]{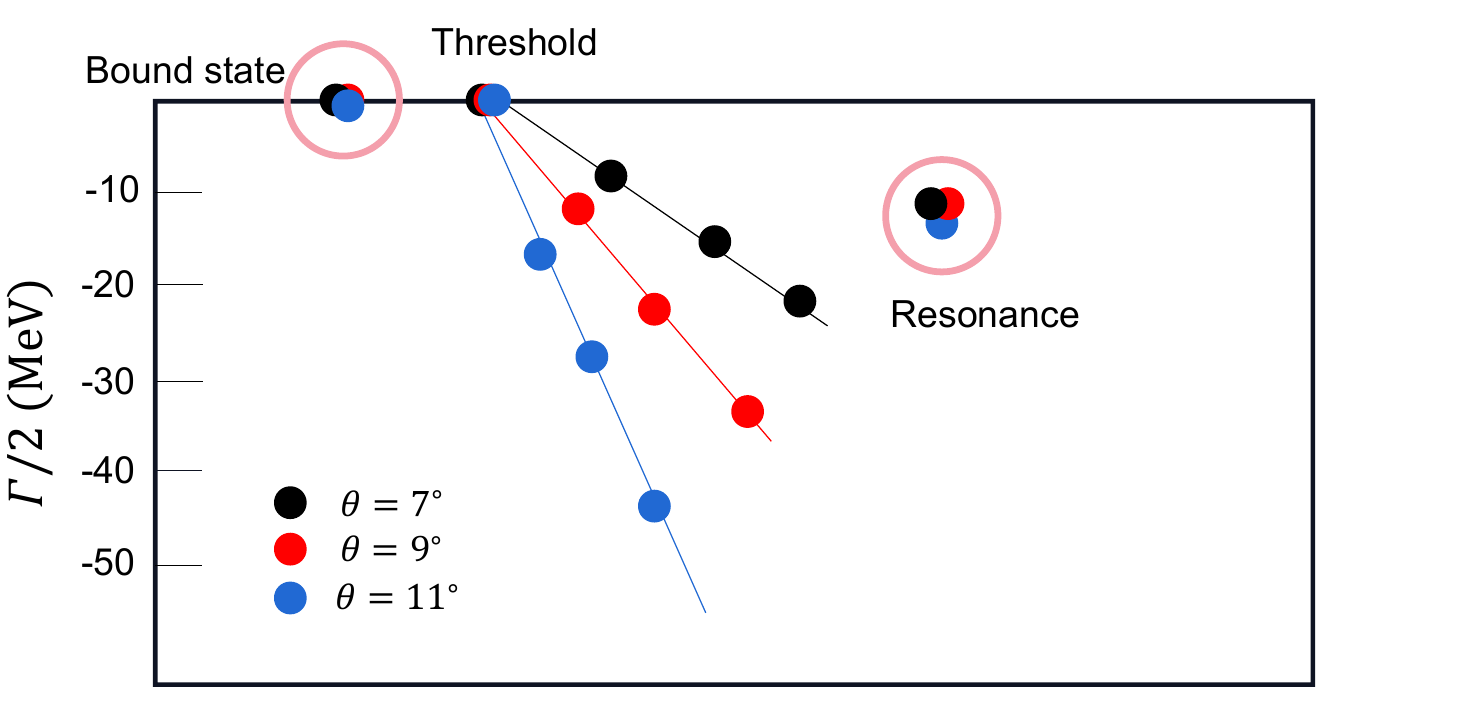}
    \caption{Schematic complex energy distribution}
     \label{csexample}
  \end{figure}

The main advantage of this method is that after transforming to the complex coordinate plane, it significantly enhances the ability to analyze resonant states, thus it provides an effective tool for investigating resonance phenomena in systems with strong interactions. A typical illustration of this process is shown in Fig. \ref{csexample}.

\section{Results and Discussions}

After all the preparation above, in this section, we aim to investigate the mass inverse problems between \( N(1440) \) and \( N(1535) \), as well as \( N(1535) \) and \( \Lambda(1405) \), from both the three-quark and five-quark perspectives. Due to the complexity of the calculations, in this work, we do not include the spin-orbit coupling in our analysis. Additionally, in the case of the five-quark calculations, it has been proposed by many researchers that \( \Lambda(1405) \) might be a \(\Sigma \pi\)-\( N\bar{K}\) two-pole structure, where the \(\Sigma \pi\) represents a bound state, and \( N\bar{K} \) corresponds to a resonance state. Therefore, we adopt the complex-scaling method to identify the possible resonant states.

\subsection{Three-Quark Calculation}

After using the model parameters listed in Table \ref{modelparameters}. The energies of the low excited $N^\ast$ and $\Lambda^\ast$ are shown in Table \ref{spectrum}. As we have mentioned in the introduction, for traditional baryon configuration on $N^\ast$ states, experimental results give that the second \( N(1/2^+) \) state, identified as \( N(1440) \), is located around 1.4 GeV, and the first \( N(1/2^-) \) state has a mass near 1535 MeV. However, our calculation gives a mass of 1767 MeV for the \( N(2S) \) state, which does not correspond to the experimental \( N(1440) \) but is closer to the experimental \( N(1710) \) state, who has a mass between 1650 and 1750 MeV. The energy of \( N(1P) \) in our calculation is 1573 MeV, which is very closed to the experimental \( N(1535) \).

\begin{table}[htbp]
\caption{Results of the hadron spectrum calculation.\label{spectrum}}
\begin{tabular}{ccccccc}
\hline\hline\noalign{\smallskip}
    Baryon                &    This work    & Ref.\cite{Leinweber:2024psf} & Ref.\cite{Zhong:2024mnt}&EXP.(PDG) \cite{ParticleDataGroup:2024cfk} \\ \hline
    $N(2S)$              &     1767        &      1900    &       1438    &             1650-1750(?)\\
    $N(1P)$              &     1573        &      -    &       1549    &             1500-1520\\
    $\Lambda(1P)$        &     1623        &      -    &       -    &             1670-1678(?)\\
\hline\hline
\end{tabular}
\end{table}

In fact, in the earlier time, many calculations on the bare three-quark state for \( N(2S) \) already predicted that its masses should be above 1.5 GeV, for example, Refs.~\cite{Capstick:1985xss,Ferretti:2011zz,Capstick:1992th}. However, it is worth noting that modifications on the quark model can lower the bare mass of \( N(2S) \) below 1.5 GeV. For example, in Ref.~\cite{Capstick:1985xss}, the authors included ad-hoc mass shifts for the \(N (1/2^-) \) and \( N(1/2^+) \) states, while Glozman et al. introduced Goldstone-boson exchange (GBE) interactions in Ref.~\cite{Glozman:1999vd}, and Yang et al. considered scalar meson exchanges in Ref.~\cite{Yang:2017qan}.  Recently, the authors in Ref.~\cite{Zhong:2024mnt} successfully reduced the energy of \( N(2S) \) to 1438 MeV and that of \( N(1P) \) to 1549 MeV by introducing pseudo-scalar meson exchange in the context of OGE potential. While these approaches address the issue of the Roper resonance mass inversion effectively, they still have some issues with other states. For example, the mass of \( \Sigma(1S) \) in Ref.~\cite{Yang:2017qan} and \( N(3S) \) in Ref.~\cite{Zhong:2024mnt} are significantly higher than their corresponding experimental values. Actually, from the experimental results of light mesons, we may do a simple estimation on the bare masses of $N(1P)$ and $N(2S)$. When we look at PDG \cite{ParticleDataGroup:2024cfk}, it is easy to see that the average mass difference between the \( 2S \)-\( 1S \) states of \( \eta \), \( \rho \), and \( \omega \) is about 700 MeV, while the mass difference for the \( 1P \)-\( 1S \) states is approximately 570 MeV. However, if we consider \( N(1440) \) as the first radial excitation of \( N(939) \), the mass difference \( 2S \)-\( 1S \) is only 500 MeV, which is significantly smaller than the \( 2S \)-\( 1S \) mass difference of mesons, while the mass difference between (\( N(1710) \)) and the ground-state \( N(939) \) is closer to the meson \( 2S \)-\( 1S \) mass difference. Therefore, we conclude that in the image of quenched quark model, the \( N(2S) \) state is unlikely to correspond to the experimental \( N(1440) \), but rather to the experimental \( N(1710) \). And such opinion that \( N(2S) \) state may not be the experimental \( N(1440) \) is also supported by the recent lattice QCD calculations \cite{Leinweber:2024psf}.

Then we look back on our calculations in Table~\ref{spectrum}. Here, the mass difference between \( N(1P) \) and \( N(1S) \) is 634 MeV, which is closer but slightly larger than the meson \( 1P \)-\( 1S \) mass difference of 570 MeV. Thus, we believe that \( N(1535) \) is likely a state primarily composed of three quarks, while \( N(1440) \) can not be a pure three-quark state. Similarly, for the $\Lambda^\ast$ family, if we treat \( \Lambda(1405) \) as the first orbital excitation of \( qqs \) state, then its mass difference from $\Lambda(1S)$ will also be much smaller than the average mass difference between the \( 1P \)-\( 1S \) states of \( \eta \), \( \rho \), \( \omega \), \( K \), and \( K^* \) mesons as approximately 510 MeV. However, this mass difference on mesons is consistent with the mass difference between \( \Lambda(1P) \) and \( \Lambda(1S) \) in our calculations. Therefore, we propose that the experimental \( \Lambda(1670) \) is likely to be the first orbital excitation of the \( \Lambda \) baryon, while \( \Lambda(1405) \) can not be a pure traditional baryon that contains three compositeness quarks.

\subsection{Five-Quark Calculation}

After studying the three-quark nature of Roper resonances, in this subsection, we perform a five-quark component discussion on \( N(1535) \) and \( \Lambda(1405) \) based on the model parameters from fitted hadron mass spectrum. For \( N(1440) \), although its mass may be around the threshold of $N\sigma$, $\sigma$ itself is so special in hadron physics that we generally do not consider coupled channels that include it. Thus, \( N(1440) \) may be unlikely to be a molecular five-quark state. Therefore, it is excluded from our discussion. Then, for the remaining states, we begin with a bound-state calculation and then use the complex-scaling method to search for possible resonance states. By using this method, we can simultaneously obtain both bound states and resonant states, along with their corresponding widths. Here, bound states are labeled as \( B(\text{energy}) \), while resonant states are labeled as \( R^{I}(\text{energy}, \text{width}) \).

\subsubsection{Bound-State Calculation}

For the five-quark components of \( N(1535) \), it is generally believed that its structure is \( qqs \)-\( \bar{s}q \). The isospin of the three-quark \( qqs \) system can take two values, \( I_{3q} = 0 \) and \( I_{3q} = 1 \), while the isospin of the two-quark \( \bar{s}q \) system is fixed at \( I_{2q} = 1/2 \). Since the total isospin of \( N(1535) \) is \( 1/2 \), it can be obtained by the couplings \( 0 \otimes 1/2 \) and \( 1 \otimes 1/2 \). Similarly, the total spin of \( 1/2 \) can arise from three possible couplings, i.e., \( 1/2 \otimes 0 \), \( 1/2 \otimes 1 \), and \( 3/2 \otimes 1 \). After considering the related symmetries, the \( qqs \)-\( \bar{s}q \) structure of \( N(1535) \) involves five channels as \( \Lambda K \), \( \Lambda K^* \), \( \Sigma K \), \( \Sigma K^* \), and \( \Sigma^* K^* \). As listed in Table~\ref{N1535}, the calculation results show that their energies range from 1.5 GeV to 2.2 GeV. Notably, the energies of \( \Sigma K \), \( \Sigma K^* \), and \( \Sigma^* K^* \) channels are below their respective threshold energies, indicating strong attractions in these channels. Finally, after performing channel coupling for all channels, the calculation results show that the lowest energy is 1569.1 MeV, which is still above the threshold of 1568 MeV, as listed in the fifth column of Table~\ref{N1535}. This suggests that our bound-state calculation does not support \( N(1535) \) as a \( \Lambda K \) molecular state.

\begin{table}[tp]
\centering
\caption{\label{N1535} The energies of the $N(1535)$  system. $i,j,k,l$ stands for the index of orbit, flavor, spin and color wave functions, respectively. $E_{th}$ means the threshold of corresponding channel, $E_{sc}$ is the energy of every single channel,  $E_{mix}$ is the lowest energy of the system by coupling all channels. (unit: MeV)}
\begin{tabular}{cccccccccc}\hline\hline
$\Psi_{\frac{1}{2},\frac{1}{2}}^{P_{ijkl}}$ ~~~ &Channel~~~           &~~~$E_{th}$ ~~~  &~~~$E_{sc}$  ~~~&~~~$E_{mix}$    \\ \hline
$\Psi_{\frac{1}{2},\frac{1}{2}}^{P_{1111}}$~~~  &$\Lambda K$          &1567.9          &1569.5         &1569.1         \\
$\Psi_{\frac{1}{2},\frac{1}{2}}^{P_{1211}}$~~~  &$\Lambda K^*$        &1986.9          &1988.2         &         &        \\
$\Psi_{\frac{1}{2},\frac{1}{2}}^{P_{1321}}$~~~  &$\Sigma K$           &1711.5          &1640.1         &        &         \\
$\Psi_{\frac{1}{2},\frac{1}{2}}^{P_{1421}}$~~~  &$\Sigma K^*$         &2130.6          &2087.5         &         &         \\
$\Psi_{\frac{1}{2},\frac{1}{2}}^{P_{1521}}$~~~  &$\Sigma^* K^*$       &2261.3          &2224.7         &         &       \\
\hline\hline
\end{tabular}
\end{table}

For \( \Lambda(1405) \), it is generally believed that it has a two-pole structure corresponding to \( \Sigma \pi \)-\( N \bar{K} \) coupling. Therefore, we consider two possible quark configurations, \( qqq \)-\( \bar{q}s \) and \( qqs \)-\( \bar{q}q \). All possible physical channels are listed in Table~\ref{L1405}, which include \( N\bar{K} \), \( N\bar{K}^{*} \), \( \Sigma \pi \), \( \Sigma \rho \), \( \Sigma^* \rho \), \( \Lambda \eta \), and \( \Lambda \omega \). The results show that the energies of the color-singlet channels are distributed between 1.4 GeV and 2.1 GeV. Among these, the \( N\bar{K} \) and \( \Sigma \pi \) channels are bound states, whose masses can correspond to the experimental states of \( \Lambda(1380) \) and \( \Lambda(1405) \), respectively. Additionally, the \( \Sigma \rho \) and \( \Sigma^* \rho \) channels also have bound states, with \( \Sigma \rho \) potentially being a candidate for the experimental state of \( \Lambda(2000) \). As for whether the \( N\bar{K} \) channel can survive in the coupled decay channels and serve as a candidate for \( \Lambda(1405) \), this is an issue that will be addressed in our subsequent resonance state calculation.

\begin{table}[tp]
\centering
\caption{\label{L1405} The energies of the $\Lambda(1405)$  system. $i,j,k,l$ stands for the index of orbit, flavor, spin and color wave functions, respectively. $E_{th}$ means the threshold of corresponding channel, $E_{sc}$ is the energy of every single channel, $E_{mix}$ is the lowest energy of the system by coupling all channels. (unit: MeV)}
\begin{tabular}{cccccccccc}\hline\hline
$\Psi_{\frac{1}{2},\frac{1}{2}}^{P_{ijkl}}$ ~~~ &Channel~~~           &~~~$E_{th}$ ~~~  &~~~$E_{sc}$  ~~~&~~~$E_{mix}$    \\ \hline
$\Psi_{\frac{1}{2},\frac{1}{2}}^{P_{1151}}$/$\Psi_{\frac{1}{2},\frac{1}{2}}^{P_{1261}}$&$N \bar{K}$            &1435.3          &1434.6        &1280.4         \\
$\Psi_{\frac{1}{2},\frac{1}{2}}^{P_{1351}}$/$\Psi_{\frac{1}{2},\frac{1}{2}}^{P_{1461}}$&$N \bar{K}^*$          &1854.3          &1854.2                 &        \\
$\Psi_{\frac{1}{2},\frac{1}{2}}^{P_{1281}}$~~~  &$\Sigma \pi$                                            &1358.5          &1317.8                 &         \\
$\Psi_{\frac{1}{2},\frac{1}{2}}^{P_{1481}}$~~~  &$\Sigma \rho$                                           &2001.8          &1983.9                &         \\
$\Psi_{\frac{1}{2},\frac{1}{2}}^{P_{1581}}$~~~  &$\Sigma^* \rho$                                         &2132.6        &2128.6               &       \\
$\Psi_{\frac{1}{2},\frac{1}{2}}^{P_{1171}}$~~~  &$\Lambda \eta$                                          &1671.2          &1672.8              &       \\
$\Psi_{\frac{1}{2},\frac{1}{2}}^{P_{1371}}$~~~  &$\Lambda \omega$                                        &1871.7        &1873.1               &       \\
\hline\hline
\end{tabular}
\end{table}

\subsubsection{Resonance State Calculation}

Then, based on the complex-scaling method with the scaling angle \( \theta \) varying from \( 7^\circ \) to \( 11^\circ \), we perform coupled-channel calculations for the five-quark systems of \( N(1535) \) and \( \Lambda(1405) \) within the GEM framework. We first perform channel coupling for all channels to identify all possible resonant states. Since our five-quark system calculations only consider molecular structure, the formation mechanism of these resonance states is derived from the binding attraction in the bound-state calculations. Therefore, based on the energies of the resonant states and their approximity to the threshold energies, the dominant components of the resonant states can be directly identified. We then couple each of these resonant states with their decay channels to explore their interactions and partial decay widths. Finally, we sum the partial widths to obtain the total decay width of each resonance state, as shown in Table~\ref{Width}.

\begin{figure}[htp]
  \setlength {\abovecaptionskip} {-0.1cm}
  \centering
  \resizebox{0.50\textwidth}{!}{\includegraphics[width=5.5cm,height=3.5cm]{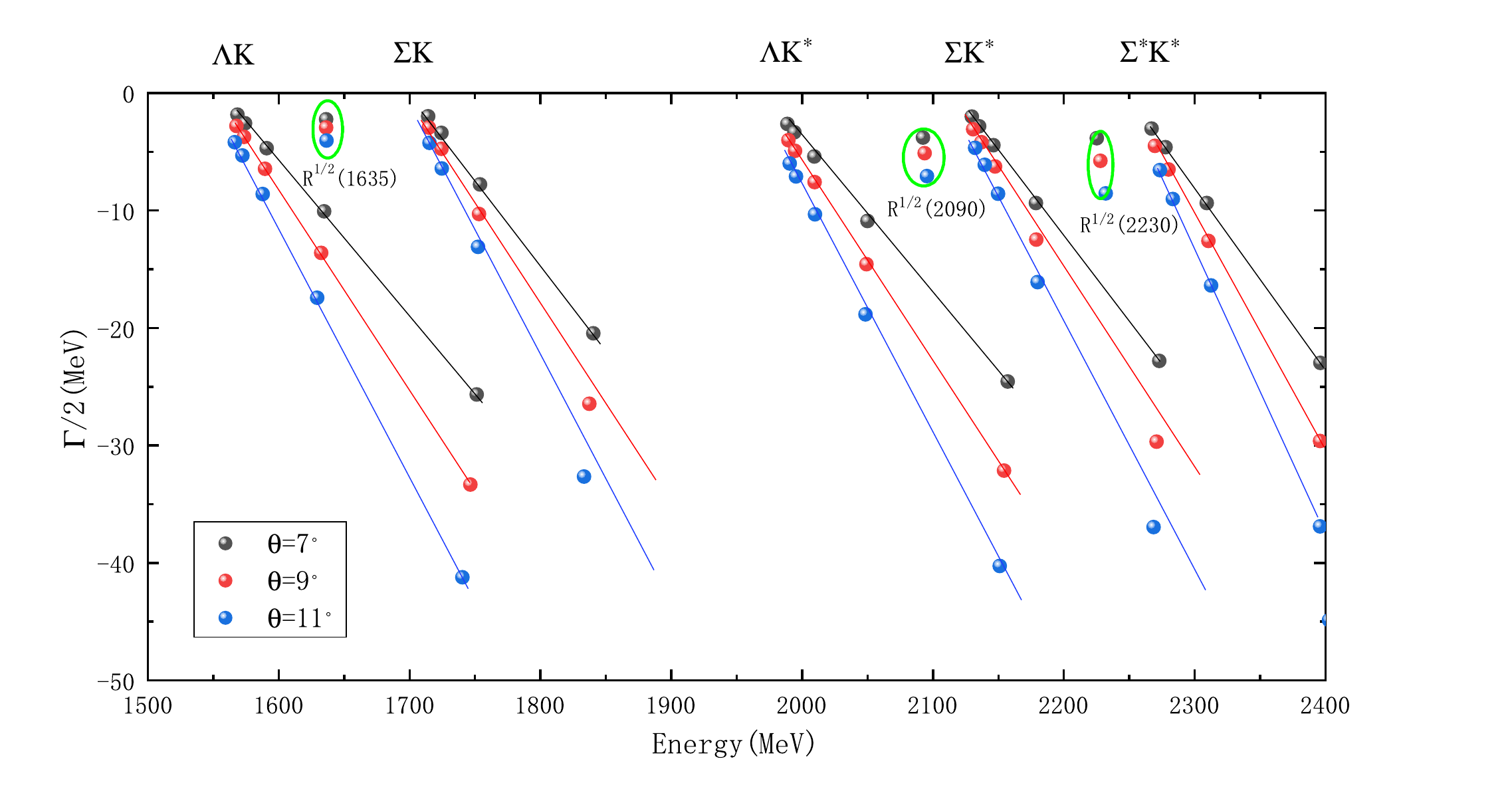}}
  \caption{ Complex-scaling results for the $N(1535)$ five-quark system in the 1500-2400 MeV range.}
\label{Fic_N}
\end{figure}

\begin{figure}[htp]
  \setlength {\abovecaptionskip} {-0.1cm}
  \centering
  \resizebox{0.50\textwidth}{!}{\includegraphics[width=5.5cm,height=3.5cm]{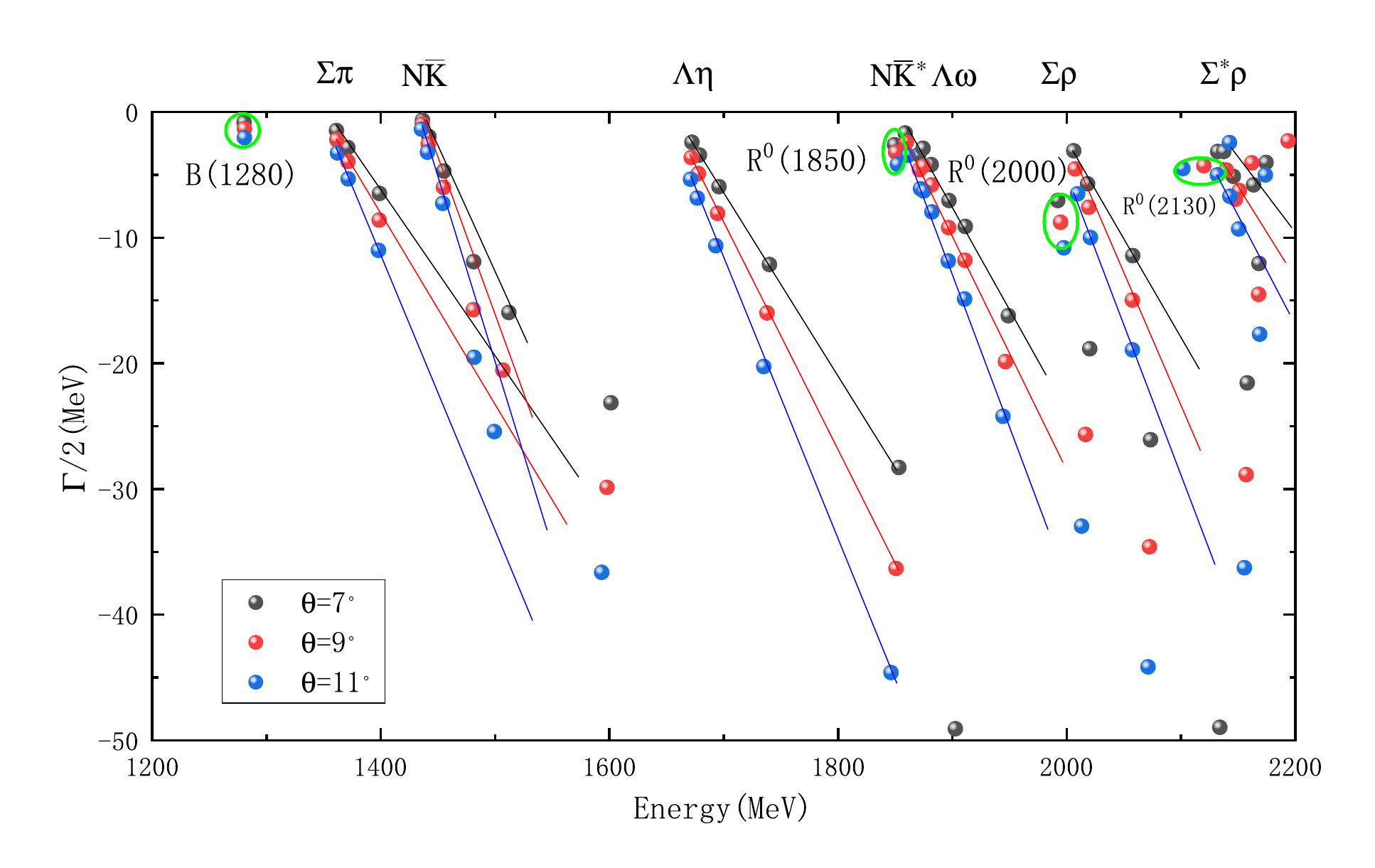}}
  \caption{ Complex-scaling results for the $\Lambda(1405)$ five-quark system in the 1200-2200 MeV range.}
\label{Fic_L}
\end{figure}

\begin{table*}[tp]
  \centering
  \fontsize{9}{8}\selectfont
  \makebox[\textwidth][c]{
   \begin{threeparttable}
   \caption{Various decay channels and corresponding decay widths of the obtained resonances. (unit: MeV)\label{Width}}
    \begin{tabular}{ccccccccc}
    \hline    \hline
   Decay channels~~~~ & $R^{1/2}(1635)$ &$R^{1/2}(2090)$&  $R^{1/2}(2230)$ &$R^{0}(1850)$ &$R^{0}(2000)$ &$R^{0}(2130)$ \\
\hline
$\Lambda K $       &  6.0          &  9.8        &  9.8           &  -        &  -        &  -    \\
$\Lambda K^* $     &  -            &  7.8        &  10.2          &  -        &  -        &  -    \\
$\Sigma K $        &  -            &  7.8        &  8.2           &  -        &  -        &  -    \\
$\Sigma K^* $      &  -            &  -          &  9.4           &  -        &  -        &  -    \\
$N K $             &  -            &  -          &  -             &  4.6      &  8.2      &  0.0    \\
$N K^* $           &  -            &  -          &  -             &  -        &  16.0     &  8.6    \\
$\Sigma \pi $      &  -            &  -          &  -             &  4.4      &  7.8      &  7.8    \\
$\Sigma \rho $     &  -            &  -          &  -             &  -        &  -        &  5.0    \\
$\Lambda \eta $    &  -            &  -          &  -             &  7.2      &  10.2     &  0.0    \\
$\Lambda \omega $  &  -            &  -          &  -             &  -        &  7.6      &  9.2    \\
Total              &  6.0          &  25.4       &  37.6          &  16.2     &  37.8     &  31.6   \\
\hline \hline
    \end{tabular}
   \end{threeparttable}}
  \end{table*}

\begin{figure*}[htbp]  
  \centering
  \setlength{\abovecaptionskip}{-0.1cm}
  \resizebox{\textwidth}{!}{  
    \includegraphics[width=12cm, height=7cm]{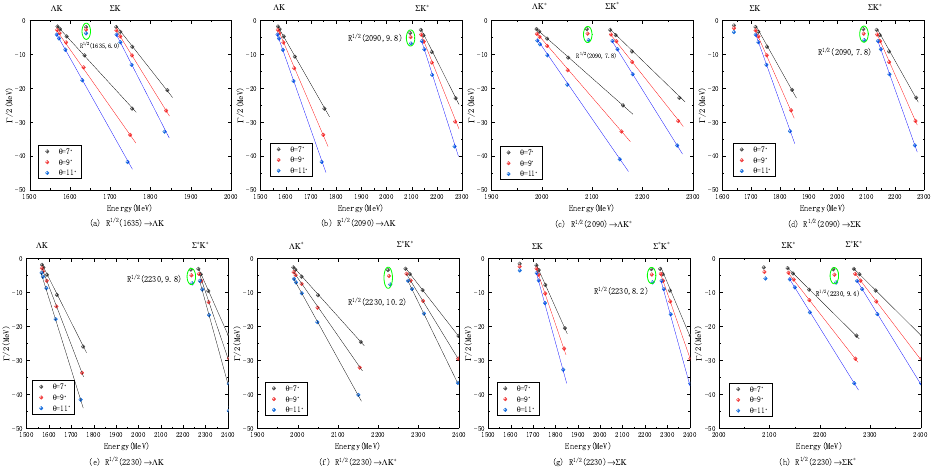} 
  }
  \caption{Calculated decay widths of resonance states in the $N(1535)$ five-quark system}
  \label{Fic_AN}
\end{figure*}

In the previous bound-state calculation on $N^{\ast}$-like states, three bound states were identified, which are \(\Sigma K\) with a binding energy of approximately 71 MeV, \(\Sigma K^*\) with a binding energy of about 42 MeV, and \(\Sigma^* K^*\) with a binding energy near 36 MeV. As illustrated in Fig.~\ref{Fic_N}, we also obtain three resonance states, one is \( R^{1/2}(1635) \), with \(\Sigma K\) as its dominant component, making it a promising candidate for the experimental \(N(1650)\). Another is \( R^{1/2}(2090) \), with \(\Sigma K^*\) as its primary component. And \( R^{1/2}(2030) \), with \(\Sigma^* K^*\) as the main component. As shown in Fig.~\ref{Fic_AN}, panel (a) presents the decay width of the resonance state \( R^{1/2}(1635) \) to \(\Lambda K\), which is around 6 MeV. As a candidate for \(N(1650)\), it is obvious that this width is much smaller than the experimental width for \(N(1650)\). This discrepancy arises because the main decay channel for \(N(1650)\) in experiments is \( N \pi \), which predominantly consists with the quark component \(qqq\)-\(\bar{q}q\), differing from our quark component \(qqs\)-\(\bar{s}q\), and thus the \(N \pi \) decay channel is not included in our calculation. In future work, we will explore the unquenched effect of introducing the five-quark structure \(qqq\)-\(\bar{q}q\). Panels (b, c, d) in Fig.~\ref{Fic_AN} display the decay widths of \( R^{1/2}(2090) \) to \(\Lambda K\), \(\Lambda K^{*}\), and \(\Sigma K\), which are 9.8 MeV, 7.8 MeV, and 7.8 MeV, respectively. Relatively, its decay width to \(\Lambda K\) is the largest, since the corresponding phase space is the largest. Panels (e, f, g, h) in Fig.~\ref{Fic_AN} show the decay widths of \( R^{1/2}(2230) \) to \(\Lambda K\), \(\Lambda K^{*}\), \(\Sigma K\), and \(\Sigma K^{*}\), with results of 9.8 MeV, 10.2 MeV, 8.2 MeV, and 9.4 MeV, respectively, for a total decay width of 37.6 MeV.

\begin{figure*}[htbp]  
  \centering
  \setlength{\abovecaptionskip}{-0.1cm}
  \resizebox{\textwidth}{!}{  
    \includegraphics[width=12cm, height=12.0cm]{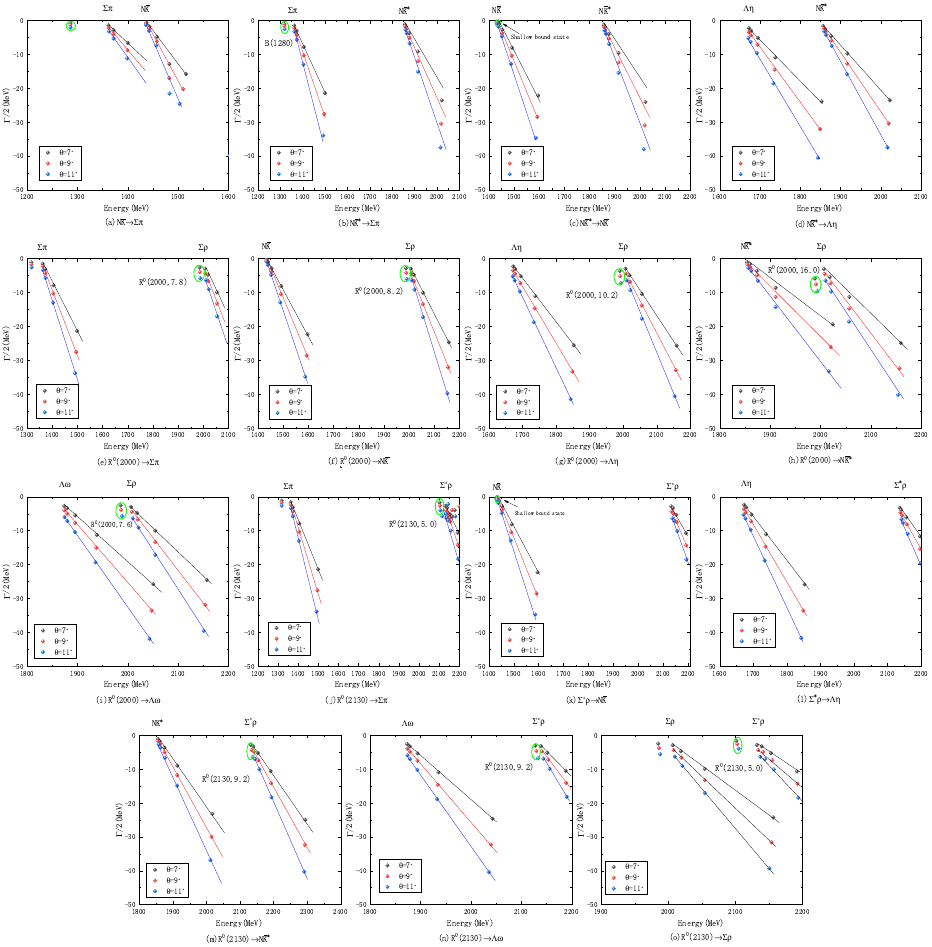} 
  }
\caption{ Calculated decay widths of resonance states in the $\Lambda(1405)$ five-quark system}\label{Fic_AL}
\end{figure*}
\begin{figure*}[htbp]  
  \centering
  \setlength{\abovecaptionskip}{-0.1cm}
  \resizebox{\textwidth}{!}{  
    \includegraphics[width=12cm, height=3.0cm]{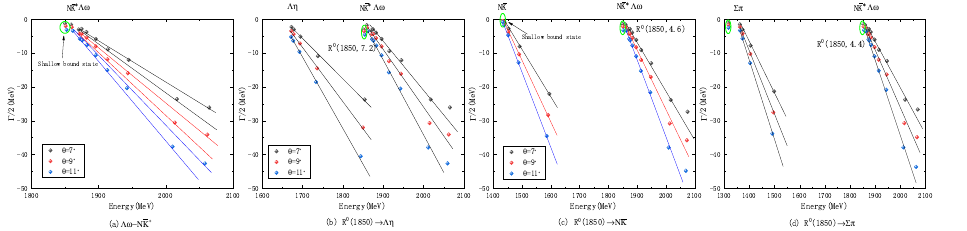} 
  }
\caption{Formation of \( R^0(1850) \) and its decay width calculation  }\label{Fic_Ap}
\end{figure*}

For \( \Lambda(1405) \), we find two deep bound states, one is \( \Sigma \pi \), which is a candidate for the experimental \( \Lambda(1380) \), the other is \( \Sigma \rho \), which is a candidate for the experimental \( \Lambda(2000) \). Additionally, three shallow bound states were obtained in \( N\bar{K} \), \( N\bar{K}^* \), and \( \Sigma^* \rho \) channels. Within the framework of the complex-scaling method, we perform a channel coupling for all channels and obtained three resonance states \( R^0(1850) \), \( R^0(2000) \), and \( R^0(2130) \), along with one bound state, \( B(1280) \), as shown in Fig.~\ref{Fic_AL}. Considering the energy positions of these states and their relations to the nearest threshold channels, the main components of these states are as follows, \( B(1280) \) is dominated by \( \Sigma \pi \), \( R^0(1850) \) is by \( N\bar{K}^* \), \( R^0(2000) \) is by \( \Sigma \rho \), and \( R^0(2130) \) is by \( \Sigma^* \rho \).  We note that the candidate for \( \Lambda(1405) \), \( N\bar{K} \), does not survive in this channel coupling. This is because the coupling between \( N\bar{K} \) and \( \Sigma \pi \) is strong enough to lower the energy of the \( \Sigma \pi \) channel by nearly 40 MeV, which in turn raises the energy of \( N\bar{K} \), turning it into a scattering state, as shown in Fig.~\ref{Fic_AL} (a). For the similar reason, the bound state \( N\bar{K}^* \) does not survive in the coupling with the decay channels, as shown in Fig.~\ref{Fic_AL} (b), (c), and (d). We guess this is because \( N\bar{K}^* \) has an energy close to \( \Lambda \omega \), making their channel coupling effect significant. Therefore, we performe a \( N\bar{K}^* \)-\( \Lambda \omega \) channel coupling calculation and coupled it with the corresponding decay channels. As shown in Fig.~\ref{Fic_Ap}, the \( N\bar{K}^* \)-\( \Lambda \omega \) coupling forms the resonance state \( R(1850) \), with a decay width of 7.2 MeV to \( \Lambda \eta \), 4.6 MeV to \( N\bar{K} \), and 4.4 MeV to \( \Sigma \pi \), resulting in a total width of 16.2 MeV. From an energy perspective, \( R^0(1850,16) \) is a candidate for the experimental \( \Lambda(1800) \). For $R^0(2130)$, the binding energy of \( \Sigma^* \rho \) is deeper than that of \( N\bar{K} \) and \( N\bar{K}^* \), thus it survives in most decay channel couplings, except for those with \( N\bar{K} \) and \( \Lambda \eta \). This suggests that the coupling effect between \( \Sigma^* \rho \) and \( N\bar{K} \), as well as \( \Sigma^* \rho \) and \( \Lambda \eta \), is relatively large.

\section{Summary}
In the framework of chiral quark model, we systematically investigated the mass inverse problem of \(N(1535)\) and \(N(1440)\) from both three-quark and five-quark perspectives, as well as the mass inversion between \(N(1535)\) and \(\Lambda(1405)\). Given the complexity of the calculations, we only consider central forces in the Hamiltonian.

The results from the three-quark calculations show that the bare mass of \(N(2S)\) is 1767 MeV, which is closer to the experimental value of \(N(1710)\). Therefore, in our quenched quark model, \(N(1440)\) cannot be explained purely as a three-quark state. In addition, considering the special role of $\sigma$, \(N(1440)\) is also unlikely to be a pure five-quark state ($N\sigma$). Then, for \(N(1P)\), its bare mass is 1573 MeV, which is in good agreement with the experimental value of \(N(1535)\). We believe that considering unquenched effects in the future will further lower the energy of \(N(1P)\), bringing it closer to the \(N(1535)\) mass. For \(\Lambda(1P)\), its bare mass is 1623 MeV, which matches the experimental value of \(\Lambda(1670)\) rather than \(\Lambda(1405)\). Therefore, in our model, \(\Lambda(1405)\) cannot be explained as a three-quark structure.

For the five-quark system, we introduce the complex-scaling method to search for possible resonance states. The results show that in the five-quark system of \(N(\frac{1}{2}^-)\), we obtain three resonances, which are \(R^{0.5}(1635,6)\), \(R^{0.5}(2090,25)\), and \(R^{0.5}(2030,38)\), with their main components being \(\Sigma K\), \(\Sigma K^*\), and \(\Sigma^* K^*\). Among these, \(R^{0.5}(1635,6)\) is a very good candidate for the experimental \(N(1650)\). For \(\Lambda(\frac{1}{2}^-)\), we obtained one bound state and three resonances, the bound state is \(B(1280)\), which can be a candidate for the experimental \(\Lambda(1380)\). One resonance is \(R^{0}(1850,16)\), which is a candidate for \(\Lambda(1800)\), another is \(R^{0}(2000,38)\), which is a candidate for \(\Lambda(2000)\), and an additional resonance \(R^{0}(2130,32)\). Although the bound-state calculation indicates that \(NK\) channel can form a bound state, it will be pushed up to a scattering state in the coupled-channel calculation. However, we believe that after introducing unquenched effects in the future, it will allow the scattering \(NK\) state to form a stable resonance state. Therefore, we propose that \(\Lambda(1405)\) is primarily a five-quark state, with three-quark component playing an important role in its resonance formation.

In summary, we conclude that the experimental \(N(1440)\) is neither a pure three-quark structure nor a five-quark structure. Thus, further study on its nature will be carried out in our next step. On the other hand, \(N(1535)\) is a good candidate for the \(N(1P)\) state. And \(\Lambda(1405)\) should be predominantly a five-quark structure, with a very important contribution from the three-quark state for its formation. In the future, unquenched effect will be introduced to investigate whether our conclusions remain consistent, with a consideration on the mixing of three-quark and five-quark states.

\acknowledgments{This work is supported partly by the National Science Foundation of China under Contract No. 12205249 and No. 12305087. Y. T. is supported by the Funding for School-Level Research Projects of Yancheng Institute of Technology under Grant No. xjr2022039. Q. H. is supported by the Start-up Funds of Nanjing Normal University under Grant No. 184080H201B20.}

\end{document}